\RequirePackage{xspace}

\def\babar{\mbox{\slshape B\kern-0.1em{\smaller A}\kern-0.1em
    B\kern-0.1em{\smaller A\kern-0.2em R}}}

\def\en         {\ensuremath{e^-}\xspace}   
\def\ep         {\ensuremath{e^+}\xspace}
 
\def\epem       {\ensuremath{e^+e^-}\xspace}

\def\mup        {\ensuremath{\mu^+}\xspace}
\def\mun        {\ensuremath{\mu^-}\xspace} 
\def\mumu       {\ensuremath{\mu^+\mu^-}\xspace}
\def\mtau       {\ensuremath{\tau}\xspace}

\def\taup       {\ensuremath{\tau^+}\xspace}
\def\taum       {\ensuremath{\tau^-}\xspace}

\def\ellell     {\ensuremath{\ell^+ \ell^-}\xspace}

\def\nub        {\ensuremath{\overline{\nu}}\xspace}
\def\nunub      {\ensuremath{\nu{\overline{\nu}}}\xspace}
\def\nub        {\ensuremath{\overline{\nu}}\xspace}
\def\nunub      {\ensuremath{\nu{\overline{\nu}}}\xspace}
\def\nue        {\ensuremath{\nu_e}\xspace}
\def\nueb       {\ensuremath{\nub_e}\xspace}

\def\num        {\ensuremath{\nu_\mu}\xspace}
\def\numb       {\ensuremath{\nub_\mu}\xspace}

\def\nut        {\ensuremath{\nu_\tau}\xspace}
\def\nutb       {\ensuremath{\nub_\tau}\xspace}

\def\nul        {\ensuremath{\nu_\ell}\xspace}
\def\nulb       {\ensuremath{\nub_\ell}\xspace}

\def\g     {\ensuremath{\gamma}\xspace}
\def\gaga  {\ensuremath{\gamma\gamma}\xspace}  
\def\ggstar{\ensuremath{\gamma\gamma^*}\xspace}

\def\piz   {\ensuremath{\pi^0}\xspace}

\def\ppz   {\ensuremath{\pi^0\pi^0}\xspace}
\def\pip   {\ensuremath{\pi^+}\xspace}
\def\pim   {\ensuremath{\pi^-}\xspace}
\def\pipi  {\ensuremath{\pi^+\pi^-}\xspace}

\def\pimp  {\ensuremath{\pi^\mp}\xspace}

\def\kaon  {\ensuremath{K}\xspace}
\def\Kbar  {\kern 0.2em\overline{\kern -0.2em K}{}\xspace}

\def\Kz    {\ensuremath{K^0}\xspace}
\def\Kzb   {\ensuremath{\Kbar^0}\xspace}
\def\KzKzb {\ensuremath{\Kz \kern -0.16em \Kzb}\xspace}
\def\Kp    {\ensuremath{K^+}\xspace}
\def\Km    {\ensuremath{K^-}\xspace}
\def\Kpm   {\ensuremath{K^\pm}\xspace}

\def\KpKm  {\ensuremath{\Kp \kern -0.16em \Km}\xspace}
\def\KS    {\ensuremath{K^0_{\scriptscriptstyle S}}\xspace} 
\def\KL    {\ensuremath{K^0_{\scriptscriptstyle L}}\xspace} 
\def\Kstarz  {\ensuremath{K^{*0}}\xspace}

\def\Kstar   {\ensuremath{K^*}\xspace}

\def\Kstarp  {\ensuremath{K^{*+}}\xspace}

\newcommand{\etapr}{\ensuremath{\eta^{\prime}}\xspace}

\def\Dbar    {\kern 0.2em\overline{\kern -0.2em D}{}\xspace}

\def\Dz      {\ensuremath{D^0}\xspace}
\def\Dzb     {\ensuremath{\Dbar^0}\xspace}
\def\DzDzb   {\ensuremath{\Dz {\kern -0.16em \Dzb}}\xspace}
\def\Dp      {\ensuremath{D^+}\xspace}
\def\Dm      {\ensuremath{D^-}\xspace}

\def\DpDm    {\ensuremath{\Dp {\kern -0.16em \Dm}}\xspace}
\def\Dstar   {\ensuremath{D^*}\xspace}

\def\Dstarz  {\ensuremath{D^{*0}}\xspace}

\def\Dstarp  {\ensuremath{D^{*+}}\xspace}
\def\Dstarm  {\ensuremath{D^{*-}}\xspace}

\newcommand{\dstr}{\ensuremath{\Dstar}\xspace}

\def\Bbar    {\kern 0.18em\overline{\kern -0.18em B}{}\xspace}

\def\BB      {\ensuremath{B\Bbar}\xspace} 
\def\Bz      {\ensuremath{B^0}\xspace}
\def\Bzb     {\ensuremath{\Bbar^0}\xspace}
\def\BzBzb   {\ensuremath{\Bz {\kern -0.16em \Bzb}}\xspace}
\def\Bu      {\ensuremath{B^+}\xspace}
\def\Bub     {\ensuremath{B^-}\xspace}

\def\Bpm     {\ensuremath{B^\pm}\xspace}

\def\BpBm    {\ensuremath{\Bu {\kern -0.16em \Bub}}\xspace}

\def\BorBbar    {\kern 0.18em\optbar{\kern -0.18em B}{}\xspace}
\def\DorDbar    {\kern 0.18em\optbar{\kern -0.18em D}{}\xspace}
\def\KorKbar    {\kern 0.18em\optbar{\kern -0.18em K}{}\xspace}

\def\jpsi     {\ensuremath{{J\mskip -3mu/\mskip -2mu\psi\mskip 2mu}}\xspace}

\mathchardef\Upsilon="7107
\def\Y#1S{\ensuremath{\Upsilon{(#1S)}}\xspace}

\def\FourS {\Y4S}

\mathchardef\Deltares="7101
\mathchardef\Xi="7104
\mathchardef\Lambda="7103
\mathchardef\Sigma="7106
\mathchardef\Omega="710A

\def\Deltabar{\kern 0.25em\overline{\kern -0.25em \Deltares}{}\xspace}
\def\Lbar{\kern 0.2em\overline{\kern -0.2em\Lambda\kern 0.05em}\kern-0.05em{}\xspace}
\def\Sigbar{\kern 0.2em\overline{\kern -0.2em \Sigma}{}\xspace}
\def\Xibar{\kern 0.2em\overline{\kern -0.2em \Xi}{}\xspace}
\def\Obar{\kern 0.2em\overline{\kern -0.2em \Omega}{}\xspace}
\def\Nbar{\kern 0.2em\overline{\kern -0.2em N}{}\xspace}
\def\Xb{\kern 0.2em\overline{\kern -0.2em X}{}\xspace}

\def\X {\ensuremath{X}\xspace}

\def\BR         {{\ensuremath{\cal B}\xspace}}

\def\bpsikl     {\ensuremath{\Bz \to \jpsi \KL}\xspace}

\def\Bzbtox     {\ensuremath{\Bzb \to \X}\xspace}

\def\taumtoe    {\ensuremath{\taum \to \en \nunub}\xspace}

\def\taumtomu   {\ensuremath{\taum \to \mun \nunub}\xspace}

\newcommand{\tev}{\ensuremath{\mathrm{\,Te\kern -0.1em V}}\xspace}
\newcommand{\gev}{\ensuremath{\mathrm{\,Ge\kern -0.1em V}}\xspace}
\newcommand{\mev}{\ensuremath{\mathrm{\,Me\kern -0.1em V}}\xspace}
\newcommand{\kev}{\ensuremath{\mathrm{\,ke\kern -0.1em V}}\xspace}
\newcommand{\ev}{\ensuremath{\mathrm{\,e\kern -0.1em V}}\xspace}
\newcommand{\gevc}{\ensuremath{{\mathrm{\,Ge\kern -0.1em V\!/}c}}\xspace}
\newcommand{\mevc}{\ensuremath{{\mathrm{\,Me\kern -0.1em V\!/}c}}\xspace}
\newcommand{\gevcc}{\ensuremath{{\mathrm{\,Ge\kern -0.1em V\!/}c^2}}\xspace}
\newcommand{\mevcc}{\ensuremath{{\mathrm{\,Me\kern -0.1em V\!/}c^2}}\xspace}

\def\mus  {\ensuremath{\rm \,\mus}\xspace}

\def\mus        {\ensuremath{\,\mu{\rm s}}\xspace}    

\def\degrees{\ensuremath{^{\circ}}\xspace}

\def\to                 {\ensuremath{\rightarrow}\xspace}

\def\pep2{PEP-II}
\def\BF{$B$ Factory}

\newcommand{\dedx}{\ensuremath{\mathrm{d}\hspace{-0.1em}E/\mathrm{d}x}\xspace}

\def\gsim{{~\raise.15em\hbox{$>$}\kern-.85em
          \lower.35em\hbox{$\sim$}~}\xspace}
\def\lsim{{~\raise.15em\hbox{$<$}\kern-.85em
          \lower.35em\hbox{$\sim$}~}\xspace}

\def\epstag  {\ensuremath{\varepsilon_{\rm tag}}\xspace}

\xspace

\newcommand{\epjBase}        {Eur.\ Phys.\ Jour.\xspace}
\newcommand{\jprlBase}       {Phys.\ Rev.\ Lett.\xspace}
\newcommand{\jprBase}        {Phys.\ Rev.\xspace}
\newcommand{\jplBase}        {Phys.\ Lett.\xspace}
\newcommand{\nimBaseA}       {Nucl.\ Instr.\ Methods Phys.\ Res., Sect.\ A\xspace}
\newcommand{\nimBaseB}       {Nucl.\ Instr.\ Methods Phys.\ Res., Sect.\ B\xspace}
\newcommand{\nimBaseC}       {Nucl.\ Instr.\ Methods Phys.\ Res., Sect.\ C\xspace}
\newcommand{\nimBaseD}       {Nucl.\ Instr.\ Methods Phys.\ Res., Sect.\ D\xspace}
\newcommand{\nimBaseE}       {Nucl.\ Instr.\ Methods Phys.\ Res., Sect.\ E\xspace}
\newcommand{\npBase}         {Nucl.\ Phys.\xspace}
\newcommand{\zpBase}         {Z.\ Phys.\xspace}

\newcommand{\nima}      [1]  {\nimBaseA~{\bf #1}}

\def\jetset74   {\mbox{\tt Jetset \hspace{-0.5em}7.\hspace{-0.2em}4}\xspace}

\def\EtaEff{(4.18 \pm 0.06\%)}
\def\NsigEta{2174 \pm 73}

\def\EtaDataChi{51/46}

\def\NqqEtaMC{371 \pm 83}

\def\tauetaBR{(1.60\pm0.05\pm0.11)\times10^{-4}}

\def\fEff{(4.08 \pm 0.07)\%}
\def\Nsigf{1255 \pm 70}

\def\fDataChi{180/152}

\def\taufBR{(3.19\pm0.18\pm0.16\pm0.99)\times10^{-4}}
\def\taufBRnofeta{(1.11\pm0.06\pm0.05)\times10^{-4}}

\def\CL{7.2\times10^{-6}}

\def\fratio{0.69 \pm0.01 \pm 0.05}

\documentclass[twocolumn,showpacs,aps,prd,superscriptaddress]{revtex4}
\usepackage{xspace}
\usepackage{graphicx}
\usepackage{dcolumn}
\usepackage{amsmath}
\usepackage{epsfig}

\newcommand{\LANLNumber} {XXXX}

\def\kk2f       {\mbox{\tt KK2F}\xspace}

\def\tt         {\taup \taum}
\def\qq         {q\overline{q}}
\def\eett       {\epem \! \rightarrow \tt}

\def\pp         { {\pim \pip} }
\def\ppp        { {\pim \pip \pim} }

\def\gg         { {\gamma \gamma} }

\def\etatogg    { \eta \! \rightarrow \gg}

\def\fone       {f_1(1285)}
\def\taufpi     {\taum \! \rightarrow  \fone \pim  \nut }
\def\feta       {\fone \! \rightarrow  \eta \pim \pip }

\def\azero      {a_0 (980)}
\def\fa         {\fone \! \rightarrow \azero \pi}

\def\taufpieta {\taum \! \rightarrow  \fone \pim\nut \rightarrow \eta \ppp \nut}

\def\taupeta     {\taum \! \rightarrow  \eta' (958) \pi^-  \nut }

\def\taupppeta     {\taum \! \rightarrow \eta \ppp \nut }

\def\tauetaprime   {\taum \! \rightarrow \eta'(958) \pim \nut}

\RequirePackage{xspace}

\usepackage{relsize}
\def\babar{\mbox{\slshape B\kern-0.1em{\smaller A}\kern-0.1em
   B\kern-0.1em{\smaller A\kern-0.2em R}}}

\begin{document}

%
\author{B.~Aubert}
\author{M.~Bona}
\author{D.~Boutigny}
\author{Y.~Karyotakis}
\author{J.~P.~Lees}
\author{V.~Poireau}
\author{X.~Prudent}
\author{V.~Tisserand}
\author{A.~Zghiche}
\affiliation{Laboratoire de Physique des Particules, IN2P3/CNRS et Universit\'e de Savoie, F-74941 Annecy-Le-Vieux, France }
\author{J.~Garra~Tico}
\author{E.~Grauges}
\affiliation{Universitat de Barcelona, Facultat de Fisica, Departament ECM, E-08028 Barcelona, Spain }
% ...
\collaboration{The \babar\ Collaboration}
\noaffiliation


\preprint{\babar-PUB-07/061} 
\preprint{SLAC-PUB-13143} 


\begin{flushleft}
 \babar-PUB-07/061\\
SLAC-PUB-13143\\
hep-ex/\LANLNumber\\[10mm]    
\end{flushleft}

{
\begin{flushleft}
\end{flushleft}

\title{
{\large \bf \boldmath   Measurement of the $\taupppeta$ Branching Fraction and a Search for a Second-Class Current in the 
$\tauetaprime$ Decay}
}

\vspace{0.5cm}

\date{\today}
\begin{abstract}
\begin{center}
\vspace{2cm}
\large \bf Abstract
\end{center}
The $\taupppeta$ decay with the $\etatogg$ mode
is studied using 384 fb$^{-1}$ of data collected by the \babar\ detector. 
The branching fraction is measured to be $\tauetaBR$. 
It is found that $\taufpieta$ is the dominant decay mode with a branching fraction of $\taufBRnofeta$.
The first error on the branching fractions is statistical and 
the second systematic. In addition, a $90\%$ confidence level upper limit on the branching fraction of the 
$\taupeta$ decay is measured to be $\CL$.
This last decay proceeds through a second-class current and is expected to be
forbidden in the limit of isospin symmetry.
\end{abstract}

\pacs{13.35.Dx, 14.60.Fg}

\maketitle

\section{Introduction}

The high-statistics sample of \mtau-pair events collected by the \babar\ experiment 
allows detailed studies of \mtau-lepton decays with small branching fractions.
Many of these decays are poorly understood and more precise measurements
of the branching fractions as well as studies of the decay mechanisms are
required.  
This work examines the $\taupppeta$ decay~\cite{chconj} where 
$\etatogg$.
We show that this mode is dominated by $\taufpi$. 
This decay mode has been previously studied by the CLEO collaboration \cite{cleo:fpi}. A measurement of the $\taufpi$ decay was also made by the \babar\ collaboration, but with a different final state \cite{babar:5prong}.

This work also presents a search for the $\tauetaprime$ decay where 
$\eta'(958)\rightarrow\eta\pim\pip$. 
Since this $\tau$ decay proceeds via a second-class current, it is expected to be forbidden in the case of
isospin symmetry. A $90\%$ confidence level upper limit has been previously set by the CLEO 
collaboration at $7.4 \times 10^{-5}$ \cite{cleo:fpi}.

This analysis is based on data recorded 
by the \babar\ detector at the \pep2\ asymmetric-energy \epem\ 
storage rings operated at the Stanford Linear Accelerator Center.
The data sample corresponds to an integrated luminosity of 384 fb$^{-1}$ recorded at
center-of-mass energies 
of 10.58 \gev and 10.54 \gev between 1999 and 2006.
With a cross section for \epem$\rightarrow \tt$ production of 
$(0.919\pm0.003)$ nb \cite{Banerjee:2007is},
this data sample contains approximately 706 million \mtau decays. 

The \babar\ detector is described in detail in Ref.~\cite{detector}.
Charged particle  momenta are measured with a five-layer
double-sided silicon vertex tracker and a 40-layer drift chamber 
inside a 1.5-T superconducting solenoidal magnet. 
A detector of internally reflected Cherenkov light (DIRC) provides $\pi/K$ separation. 
A calorimeter consisting of CsI (Tl) 
crystals is used to measure the energy of electromagnetic showers,
and an instrumented magnetic flux return (IFR) is used to
identify muons.

Monte Carlo simulation is used to evaluate the 
background contamination and selection efficiencies.
The \mtau-pair production is simulated with the KK2F Monte Carlo event
generator \cite{kk}. The \mtau decays, continuum $q\bar{q}$ events, and final-state radiative effects are modeled
with Tauola \cite{tauola1}, JETSET \cite{qq1}, and Photos \cite{tauola2}, respectively. 
The generic $\tau$ Monte Carlo sample contains one decay mode with an $\eta$ meson in the final state, $\tau^- \rightarrow \eta \pi^- \pi^0 \nu_\tau$. In addition, a dedicated Monte Carlo sample is generated using KK2F and Tauola for the $\tau^-\rightarrow\eta K^0_s\pi^-\nu_{\tau}$ decay. 

Dedicated samples of $\tt$ events are created 
using EvtGen \cite{evtgen} where
one of the $\tau$ leptons can decay to any mode included in Tauola \cite{tauola1} and the other $\tau$ 
decays to  an $\eta\pi^-\pi^+\pi^-\nu_{\tau}$ final state. 
One of the samples is generated using the $\taufpi$ decay, and the other is generated using $\taupppeta$ phase space. 
The $\fone$ meson decay modes that are relevant to this analysis are the 
$\feta$ (40\%) and the $\fa$ (60\%) decay modes where
the relative contributions to the $\taufpi$ decay are indicated in parentheses \cite{PDG}. The Monte Carlo distributions identified as signal in the figures use a combination of $\taupppeta$ phase space, and the resonant $\taufpi$ decay samples where the relative fraction is based on the branching fractions measured in this work. 

The detector response is simulated with GEANT4 \cite{GEANT}. All Monte Carlo simulation events are passed through a full simulation of the \babar\ detector and are reconstructed in the same way as the data \cite{detector}.

\section{Selection}
Events of interest are isolated with a loose pre-selection. 
Since \mtau pairs are produced back-to-back in the \epem center-of-mass 
frame, the event is divided into hemispheres according to the thrust axis \cite{thrust}, 
calculated using all reconstructed charged and neutral particles. 
The analysis procedure selects events with one track in one 
hemisphere (tag hemisphere) and three tracks in the other hemisphere
(signal hemisphere).
The total event charge is required to be zero.
  
Charged particles are required to have transverse momenta greater 
than $0.1\gevc$ in the laboratory frame.
The distance of the point of closest approach of the track to the 
beam axis must be less than 1.5 cm.  
In addition, the $z$ coordinate (along the beam axis) of the 
point of closest approach of the track must be 
within 10 cm of the $z$ coordinate of the production point. 
Neutral clusters are required to have an energy of at least 30 \mev and must not be associated with a charged track.

After pre-selection a more discriminating analysis selection is applied. This selection strategy has three components. The first selection criterion is 
based on the event-shape properties. The magnitude of the thrust is required to be between 0.92
and 0.99, in order to reduce the non-$\tau$ backgrounds. 
The background from non-$\tau$ sources can arise from Bhabha, di-muon,
two-photon and $q\bar{q}$ events. 

The second set of selection criteria requires the particles
in the tag hemisphere to originate  from a leptonic $\tau$ decay ($\taumtoe$ or $\taumtomu$).
The track in the tag hemisphere must be identified as an electron 
or muon and must have a momentum in the center-of-mass frame below 
$4\gevc$.
The first criterion removes $q\bar{q}$ events while the second removes
lepton-pair events.
Electrons are identified with the use of the ratio of energy measured by 
the calorimeter to track momentum $(E/p)$, the ionization loss in the 
tracking system  $(\dedx)$, and the shape of the shower in the calorimeter.
Muons are identified by hits in the IFR and energy deposits in the
calorimeter expected for a minimum-ionizing particle. 
Residual background from $q\bar{q}$ events is reduced by requiring 
that there be at most one electromagnetic calorimeter cluster 
in the tag hemisphere with energy above 50 \mev and that 
the total neutral energy in the tag hemisphere be
less than 1 \gev.
 
The final set of selection criteria is applied to the signal hemisphere. The aim is to reduce the residual 
backgrounds from $\tau$ decays while maintaining high selection efficiency for $\taupppeta$ decays. An event is rejected if any of the tracks in the signal hemisphere 
is identified as an 
electron or if any pair of oppositely charged tracks is consistent 
with originating from a photon conversion.    

The event selection requires that there be one  unique $\etatogg$ candidate in the signal 
hemisphere. The $\etatogg$ candidates consist of two neutral clusters in the 
electromagnetic calorimeter with an invariant mass ($M_\gg$) 
between 0.47 and 0.62 \gevcc. To reduce combinatoric background from other $\tau$ decays with $\pi^0$ mesons, the higher
and lower-energy clusters must have $E>0.7$ GeV and $E>0.3$ GeV, respectively.

Residual background from other $\tau$ decays and $q\bar{q}$ events is
reduced by requiring that there be no $\pi^0$ mesons in the signal
hemisphere, where a $\pi^0$ candidate consists of two neutral clusters in the electromagnetic calorimeter with an invariant $\gamma\gamma$ mass between 115 and 150 \mevcc. 
In addition the invariant mass of the  $\eta \ppp$ system is required to be less than $1.8 \gevcc$. 
No particle identification algorithm is applied to the charged tracks to distinguish pions from kaons, and all invariant masses are calculated assuming that the charged particles 
are pions. Note that the $\taum \! \rightarrow  \fone \Km  \nut$ decay is kinematically disfavored. Monte Carlo studies indicate that the background from $\tau$ decays is dominated by $\gamma\gamma$ combinatorics in the $\pi^-\pi^+\pi^-(\geq1\pi^0)$ mode. The absolute amount of $\tau$ background is determined in the fits. The backgrounds from $\tau^-\rightarrow\eta \KS\pi^-\nu_{\tau}$ and $q\bar{q}$ are evaluated separately and discussed in section III.

\begin{figure}[ht]
\begin{center}
\includegraphics[height=5cm]{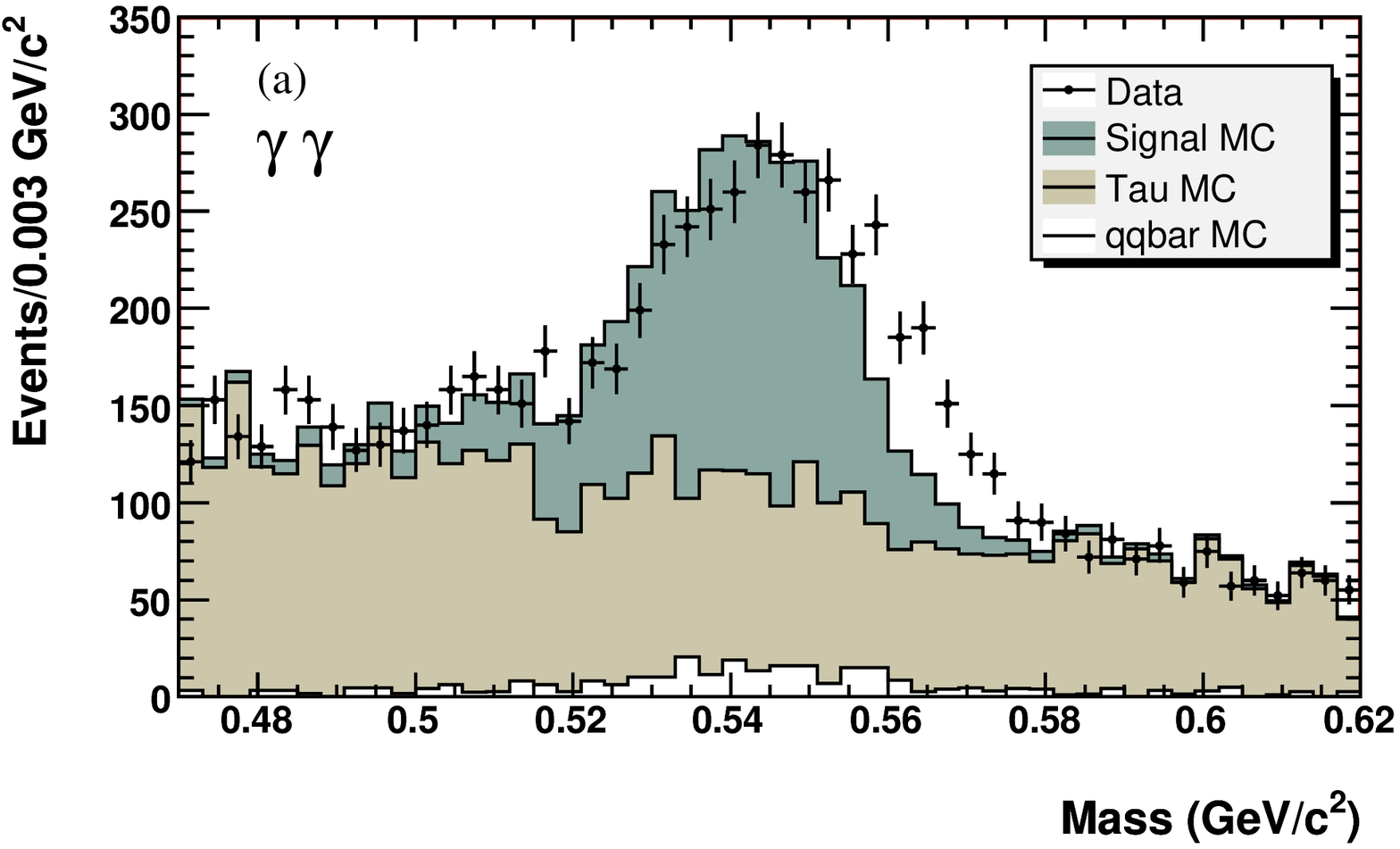}
\includegraphics[height=5cm]{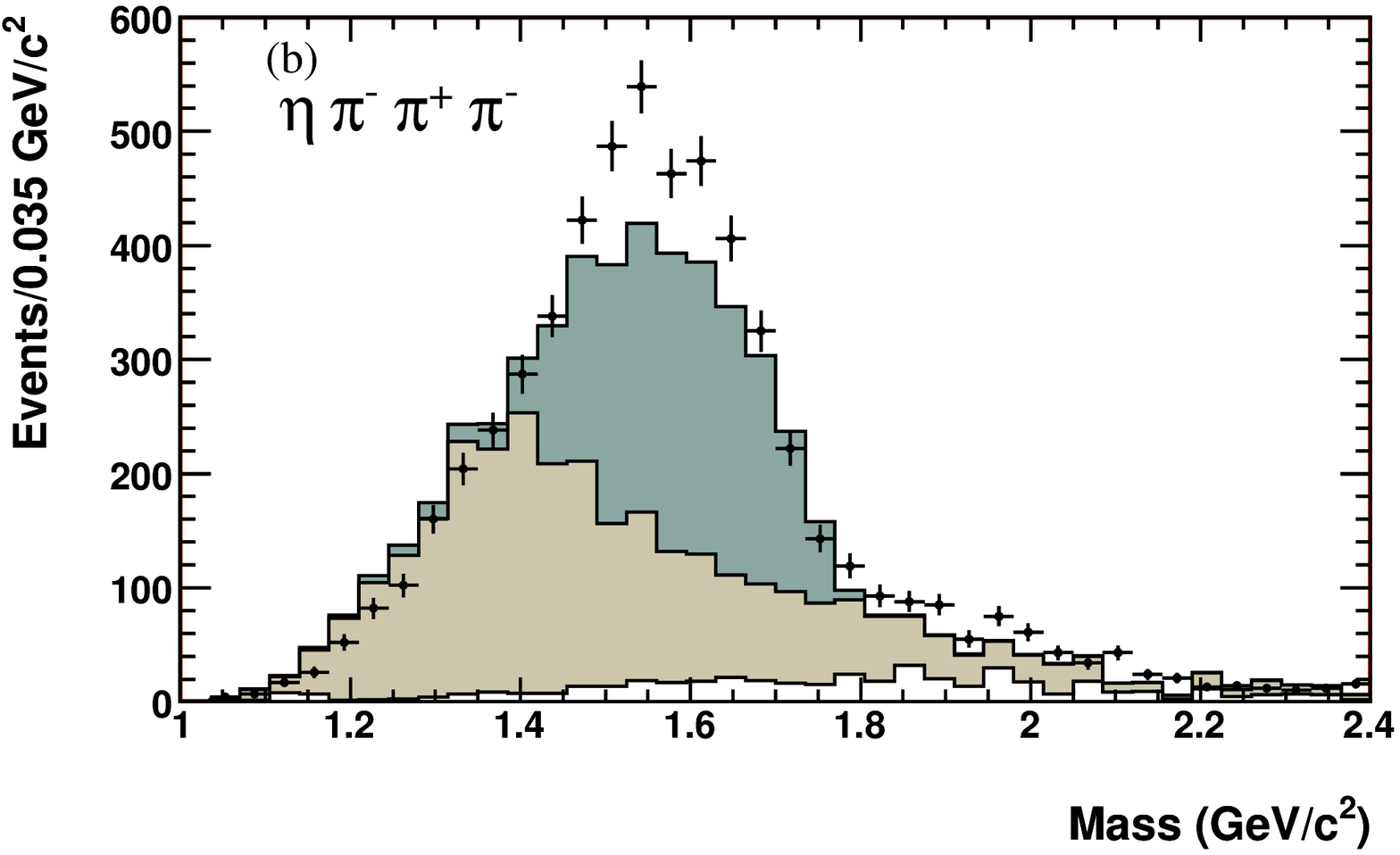}
\end{center}
\caption{\label{plot:one}
The invariant masses of the (a) $\gg$, and (b) $\eta \ppp$
final states are shown.
The dark shaded histograms show the simulated signal events, 
the lightly shaded histograms show the simulated $\tau$ background and 
the unshaded histograms show the simulated $\qq$ background.
All selection criteria are applied.
In (b) the cut requirement on the invariant mass of the $\eta \ppp$ 
system is not imposed and the invariant mass of the
$\gamma \gamma$ system is between 0.50 and 0.58 \gevcc. 
}
\end{figure}

\begin{figure}[ht]
\begin{center}
\includegraphics[height=5cm]{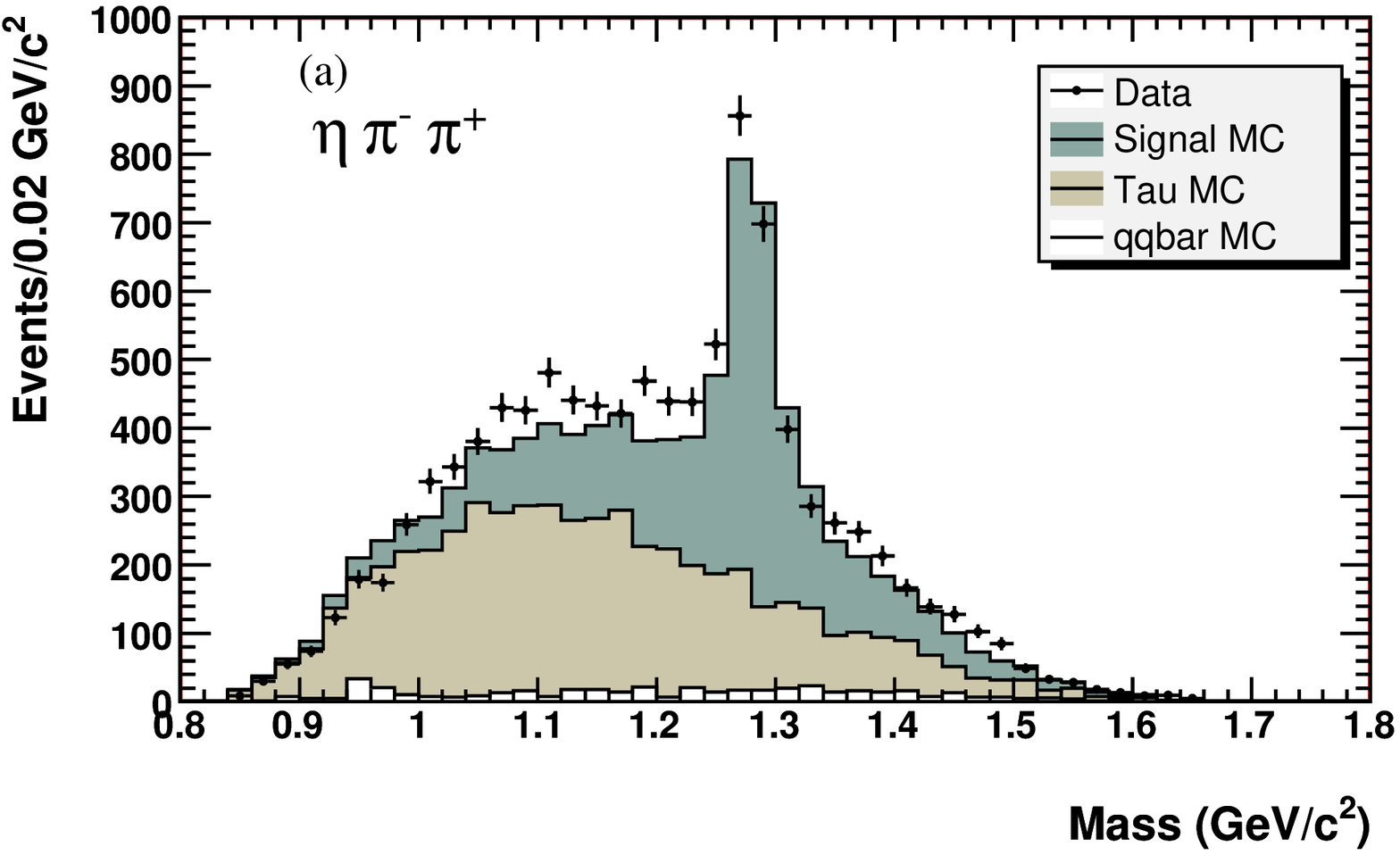}
\includegraphics[height=5cm]{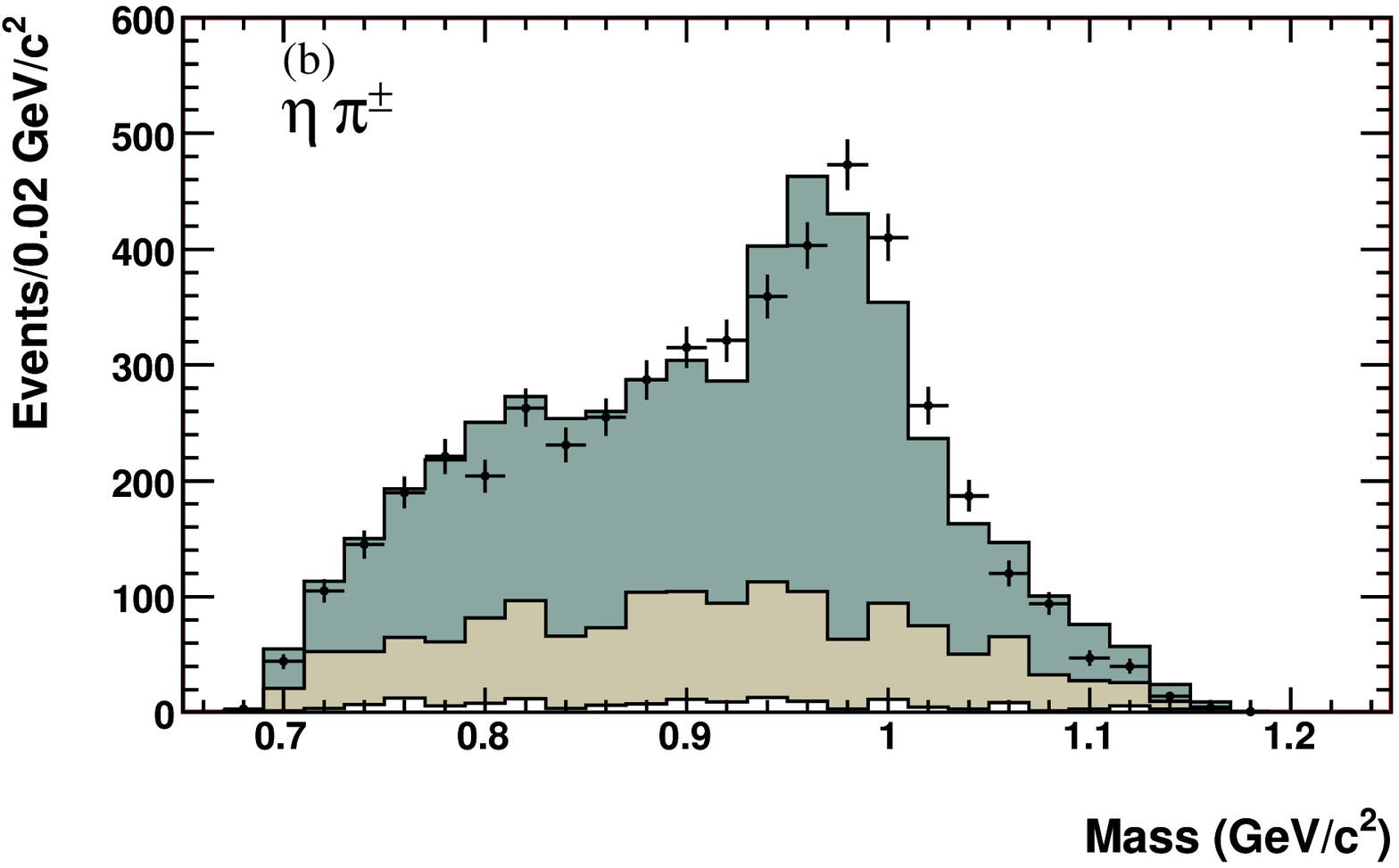}
\end{center}
\caption{\label{plot:two}
The invariant masses of the (a) $\eta \pp $ and 
(b) $\eta \pi^{\pm}$ final states are shown.
The dark shaded histograms show the simulated signal events, 
the lightly shaded histograms show the simulated $\tau$ background and 
the unshaded histograms show the simulated $q\bar{q}$ background.
Note that (a) and (b)  have two entries per event.
All selection criteria are applied. 
In (b), it is required that the invariant mass of 
the associated $\eta \pp$  system is between 1.23 and 1.32 \gevcc.
}
\end{figure}

\section{Results}

The invariant mass of the $\eta$ candidates is shown in 
Fig.~\ref{plot:one}(a) after all selection criteria are applied.
The plot shows a difference in the reconstructed mass and width of the $\eta$ meson between the data
and Monte Carlo simulation. The width of the $\eta$ peak obtained in a fit using only the signal Monte Carlo sample
was slightly narrower but consistent with the width from the data.  The variations in the fit results between 
data and signal Monte Carlo are accounted for with a systematic error discussed in the following section. 

Figure ~\ref{plot:one}(b) shows the invariant mass of the $\eta \ppp $ 
system after all selection criteria (except on this variable) have 
been applied. The disagreement between the data and Monte Carlo simulation in Fig.~\ref{plot:one}(b) shows that the underlying physics is more complex than the model used to simulate the decay and may involve additional resonances. For example the model used in Tauola for the $\tau\rightarrow 3\pi 2\pi^0\nu_{\tau}$ mode, does not give an accurate representation of the experimental data \cite{Sobie:2005sq}. However, the $\eta \ppp $ invariant mass distribution is not used in the determination of the branching fractions presented in this paper and the modeling uncertainties are small and included in the systematic errors. 

Figure~\ref{plot:two}(a) shows the invariant mass of the $\eta \pp$ system. 
Figure~\ref{plot:two}(b) shows the invariant mass of the $\eta \pi^-$ 
and $\eta \pi^+$ systems with the requirement that the invariant mass of the
$\eta \pp $ system is between 1.23 and 1.32 \gevcc 
(consistent with being an $\fone$ meson).
Only the $\pi$ mesons forming the $\fone$ candidates are shown in this plot.
The peak at 980 \mevcc is due to the $a_0(980)$ in the $\fa$ decay.

\subsection{\boldmath{Inclusive $\mathbf{\taupppeta}$ branching fraction}}

The invariant mass distribution of the $\gg$ system is fitted with a 
Novosibirsk function \cite{novo} (Gaussian distribution with a tail parameter)
for the $\eta$ meson plus a polynomial function for the background. 
The fit range is 0.47 to 0.62 \gevcc and the fit is a binned $\chi^2$ fit.
The observed width of the $\eta$ is dominated by the experimental resolution
(14 \mevcc).
The resolution and the tail parameters in the fit to the data
are fixed to the values obtained from a fit to the signal Monte Carlo simulation 
(see the following paragraphs for a discussion of the systematic error associated with this constraint).
The peak position and normalization parameters of the Novosibirsk function are 
allowed to vary in the fit to minimize the dependence of the result
on the difference in the $\eta$ mass observed between
the data and Monte Carlo simulation.

\begin{figure}[!htb]
\begin{center}
\includegraphics[height=5cm]{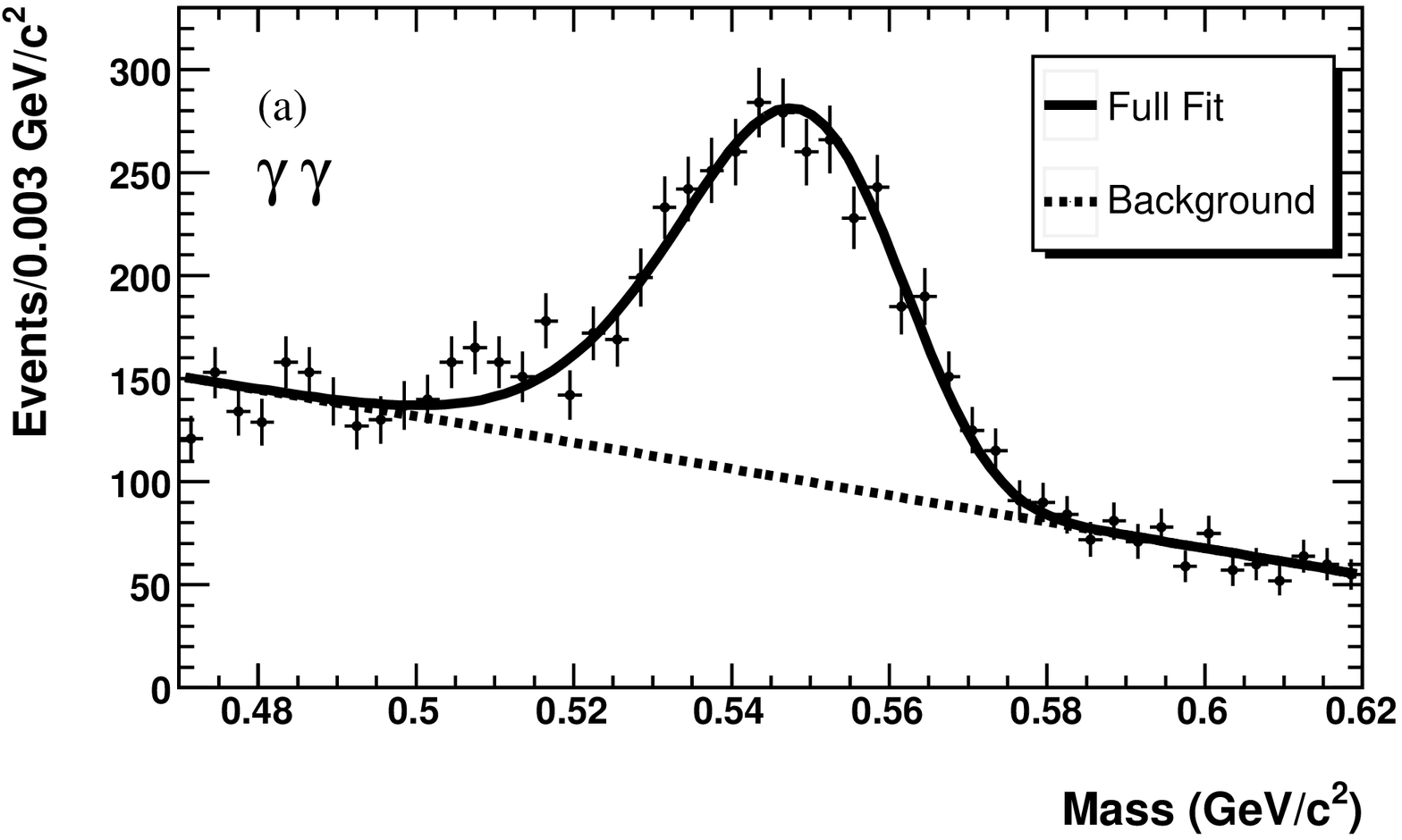} 
\includegraphics[height=5cm]{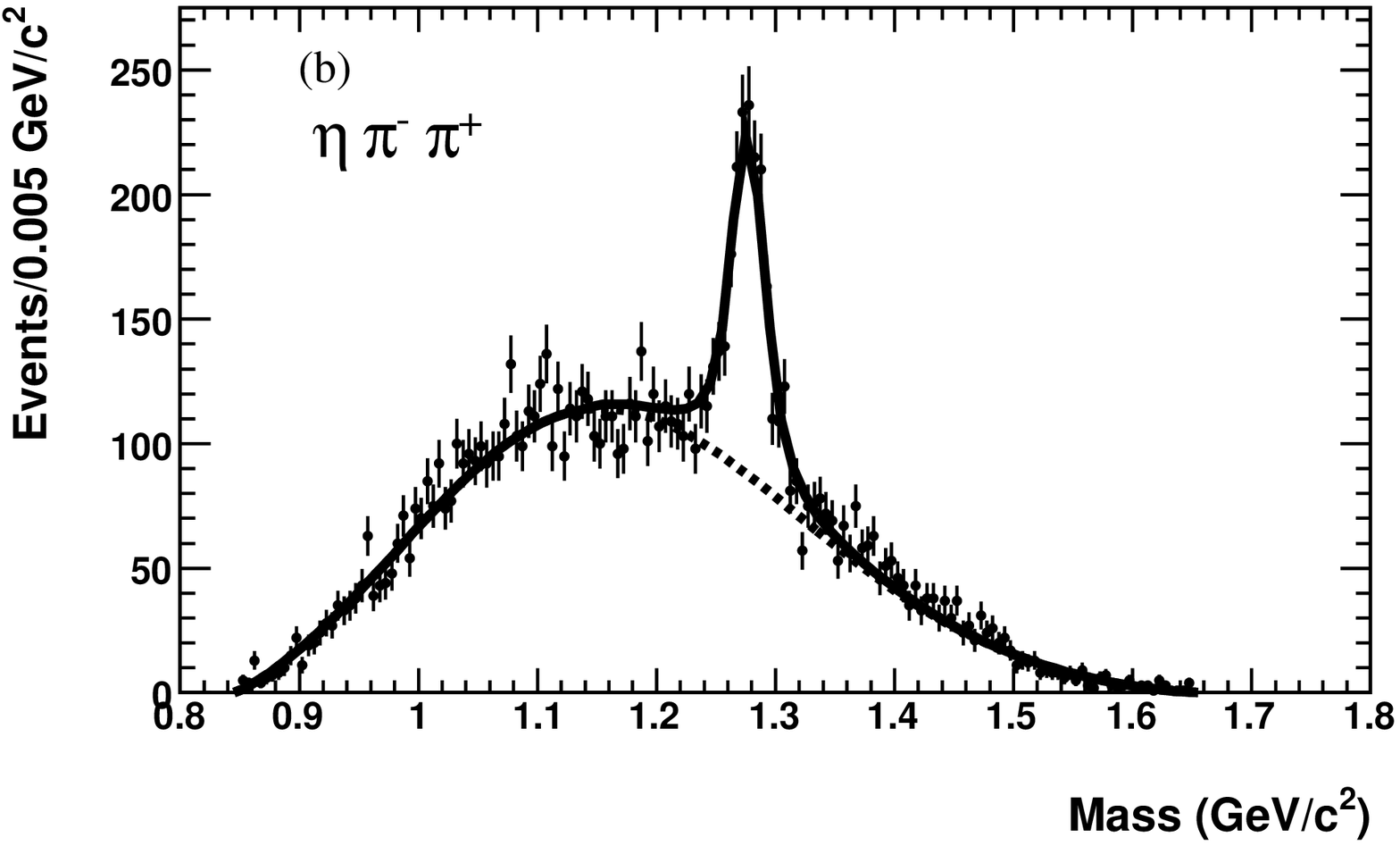}
\end{center}
\caption{\label{plot:eta-fit}
Plot (a) shows the invariant mass of the $\gg$ system. 
Plot (b) shows the invariant mass of the data $\eta \pp$ system. 
The points represent the data, the solid line
is the fit function, and the dashed line shows the background.
}
\end{figure}

A total of $\NsigEta$ events are obtained from the fit shown 
in Fig.~\ref{plot:eta-fit}(a). The $\chi^2/\mathrm{n.d.f}$ for the fit is $\EtaDataChi$.

The $\taupppeta$ branching fraction is measured with 
\begin{equation}
B_{\taupppeta} = \frac{N_{obs} - N_{bkgd}}{2 N_{\tt}} \frac{1}{\epsilon} 
\frac{1}{B(\etatogg)},
\end{equation}
where $N_{obs}$ is the number of events obtained from the fit,
$N_{bkgd}$ is the number of background events with an $\eta$ meson 
($\NqqEtaMC$), $N_{\tt}$ is the number of $\tau$ leptons in the sample calculated from 
the luminosity and $\eett$ cross section,
$\epsilon$ is the efficiency for selecting the signal events $\EtaEff$, 
and $B(\etatogg)$ is $0.3943 \pm 0.0026$ \cite{PDG}.
The $\taupppeta$ branching fraction is measured to be $\tauetaBR$
where the  first error is statistical and the second is 
systematic.

The systematic errors are dominated by the uncertainty in the 
results of the fit to the $\eta$ meson (5.0\%), which is partly due to the difference in the $\gamma\gamma$ mass resolutions in the data and Monte Carlo. 
The sensitivity of the results to the fit to the $\eta$ peak is investigated 
by unconstraining the width and tail parameters in the fit to the data.  
Also, polynomials of different orders are tested as background functions.
The 5.0\% uncertainty associated with the fit is also due partly 
to the variation of the branching fraction observed for different
background functions. 
The remaining systematic errors include terms for the uncertainties of the 
$\eta$ background levels (3.8\%),
$\eta$ selection efficiency (3.0\%),
track reconstruction (2.4\%), 
lepton identification (1.6\%), 
selection efficiency statistical error (1.4\%), 
luminosity (1.0\%), 
and the $\etatogg$ branching fraction (0.7\%).

The background events ($N_{bkgd}$) includes a contribution from $q\bar{q}$ events which is estimated from $q\bar{q}$ Monte Carlo samples. The uncertainty in the number of background events extracted from the $q\bar{q}$ background is evaluated by comparing data and Monte Carlo simulation distributions in regions where 
there is an enhanced amount of $q\bar{q}$ events (events with
an  $\eta \ppp$ invariant mass that is larger than the $\tau$ mass). The $q\bar{q}$ Monte Carlo predicts $125$ candidates with an $\eta$ meson with an uncertainty of 18 events. 

The background also gets a contribution from $\tau^-\rightarrow\eta \KS\pi^-\nu_{\tau}$ events.
The number of $\KS$ candidates is determined by counting the number of events that pass the full selection from the dedicated $\tau^-\rightarrow\eta \KS\pi^-\nu_{\tau}$ Monte Carlo. The Monte Carlo predicts $246$ $\KS$ background events. The uncertainty on the number of selected background events is dominated by the uncertainty of the $\tau^-\rightarrow\eta \KS\pi^-\nu_{\tau}$ branching fraction $B_{\tau^-\rightarrow\eta \KS\pi^-\nu_{\tau}} = (1.10\pm0.35\pm0.11)\times10^{-4}$ \cite{Bishai:1998gf}. The total background is estimated to be $N_{bkgd} = 371\pm83$ where the uncertainties from the $q\bar{q}$ and $\KS$ backgrounds are added in quadrature and included as a systematic error. 

The stability of the branching fraction measurements was tested by varying the selection criteria (within a range of values determined by the level of agreement between data and Monte Carlo), which
did not change the results significantly.  
Furthermore, the results of the fit to the $\eta$ meson mass peak are found to be 
stable to variations in the bin width or mass range used in the fit.
Branching fractions measured on the electron and muon samples separately are found to be consistent.

\subsection{\boldmath{Branching fractions for $\taufpieta$  and $\taufpi$}}

The $\fone \pi$ candidates are selected with the criteria specified in the previous section with the requirement 
that the $\eta$ candidates have $0.50 < M_\gg < 0.58$ \gevcc.
The invariant mass of the $\eta \pim \pip $ system is fitted with a Breit-Wigner function 
convoluted with a Gaussian distribution
and the background is modeled with a Novosibirsk function summed with a polynomial 
(see Fig.~\ref{plot:eta-fit} (b)). The $\chi^2/\mathrm{n.d.f.}$ is $\fDataChi$ for the $\eta \pi^-\pi^+$ fit. A P-wave Breit-Wigner function \cite{pvals} is used to fit the data while the fit to the simulated distribution uses a simple Breit-Wigner function as implemented in the generator. In both cases the Breit-Wigner function is modulated by phase space. 
The normalization and mean of the Breit-Wigner function
are allowed to vary and the width is fixed to the PDG value.   
The resolution parameter of the Gaussian function is fixed to 7 \mevcc, which is the mass resolution of the $\eta \pi^-\pi^+$ system obtained from simulation.

The background function is determined by fitting the $\eta\pi^-\pi^+$ invariant mass distribution obtained from a sample of simulated $\taupppeta$ events where the decay does not proceed through an $f_1(1285)$ meson.

There is no evidence for the production of $\fone$ mesons in the data from 
background sources. This is determined by relaxing selection criteria to increase the background from multihadron events.

The $\taufpieta$ branching fraction is determined with  
\begin{equation} 
B_{\taupppeta (\mathrm{via \hspace{.1cm}  f_1})} = \frac{N_{obs}}{2 N_{\tt}} 
\frac{1}{\epsilon} \frac{1}{B(\etatogg)}Ã,Â 
\end{equation}
where $N_{obs}$ is the number of $\fone$ mesons obtained in the fit ($\Nsigf
$),
$N_{\tt}$ is the number of $\tau$ pairs obtained from the luminosity
and $\eett$ cross section,
$\epsilon$ is the efficiency for selecting a $\taufpi$ event $\fEff$, and
$B(\etatogg)$ is the $\etatogg$ branching fraction ($0.3943 \pm 0.0026$) \cite{PDG}. 
The $\taufpi$ branching fraction is determined by dividing Eq. 2 by the $\feta$ rate ($0.35 \pm 0.11$) \cite{PDG}.

The branching fractions 
for the  $\taufpieta$ and $\taufpi$ modes are $\taufBRnofeta$ and $\taufBR$ respectively,
where the first error is statistical and the second is systematic.
The third error quoted on the $\taufpi$ measurement is due to the 
large error on the $\feta$ branching fraction.
Most systematic errors for these branching fractions are common 
to the ones listed for the inclusive measurement. While the $\eta$ fit uncertainty affects the inclusive result only, 
an extra systematic error of 1\% comes through the $\fone$ decay modeling due to the uncertainty 
of the branching fractions of the $\fa$ and $\feta$ decay modes \cite{PDG}.
This is determined by varying the relative contribution of the two modes
within the quoted uncertainties.

The fraction of the $\taufpieta$ mode 
to the inclusive $\taupppeta$ mode is found to be
$\fratio$ where the first error 
is statistical and the second is systematic (taking into account the
correlations between the various components).

\subsection{\boldmath{Limit on the $\mathbf{\taupeta}$ branching fraction}}

A limit on the $\taupeta$ branching fraction can be set by
searching for decays of the $\eta'(958)$ to the $\eta \pp $ final state.
This $\tau$ decay mode proceeds through a forbidden second-class current and is not 
expected to produce an observable signal \cite{Neufeld:1994eg}.

\begin{figure}[!htb]
\begin{center}
\includegraphics[height=5cm]{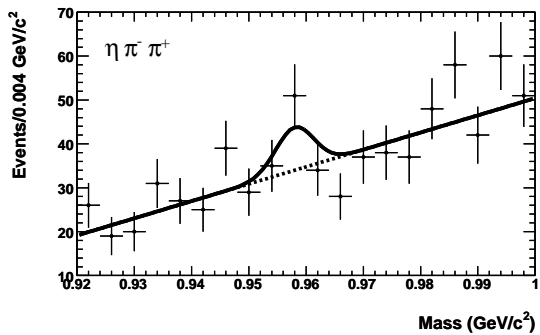}
\end{center}
\caption{\label{plot:eta'-fit}
Fit to the $\eta'$(958) region of the $\eta\pi^-\pi^+$ invariant mass spectrum. The fit uses a Gaussian distribution for the peak summed with a linear function for the background. 
The events are tagged with muons or electrons, and all selection criteria have been applied.
}
\end{figure}

A fit to the $\eta\pi^-\pi^+$ mass distribution is performed with a Gaussian function for the $\eta'(958)$ and a 
polynomial function for the background (see Fig.~\ref{plot:eta'-fit}).  
The mean of the Gaussian is fixed to the mass of the $\eta'(958)$ meson.
The width of the Gaussian distribution is fixed to the value 
obtained in a fit to a data sample containing a significant number of $\eta'(958)$ mesons.
This data sample is created by removing all selection criteria except the loose pre-selection described in section II.

We observe $19 \pm 13$ candidates.
To set a limit, we treat all of the events
in the $\eta'(958)$ peak as signal. We assume that the efficiency for selecting $\taupeta$ events is
the same as the $\taufpi$ selection efficiency.
The systematic uncertainty is dominated by a 
7\% error due to the uncertainty in the mass resolution of the $\eta'(958)$.
The remaining systematic errors are the same as those described in the previous section.
The results give a 90\% confidence level upper limit on the 
$\taupeta$ branching fraction of $\CL$.

\section{Summary}
The $\taupppeta$ decay using the $\etatogg$ mode 
is studied with the \babar\ detector.  
It is found that $\taufpi$ is the dominant decay mode for the 
$\eta \ppp $ final state.

The branching fraction of $\taupppeta$ is measured to be $\tauetaBR$
where the first error is statistical and the second systematic.
This measurement is more precise than the CLEO result $(2.3\pm 0.5) \times 10^{-4}$ \cite{Anastassov:2000xu}.

The branching fraction of the $\taufpieta$ decay mode is 
measured to be $\taufBRnofeta$ and is consistent with previous results \cite{PDG}.

The branching fraction of   
$\taufpi$ is measured to be $\taufBR$. The first error 
is statistical, the second is systematic, and the third error is associated with the $30\%$ uncertainty on the 
$f_1\rightarrow\eta\pi^-\pi^+$ branching fraction \cite{PDG}. 
This measurement is in agreement with the CLEO result of
$5.8^{+1.4}_{-1.3}\times 10^{-4}$ \cite{cleo:fpi} and the \babar\ result 
of $(3.9 \pm 0.7 \pm 0.5)\times 10^{-4}$ \cite{babar:5prong}. The branching fraction
of $\tau^- \rightarrow f_1(1285) \pi^- \nu_{\tau}$ is predicted by effective chiral theory
to be $2.9\times10^{-4}$ \cite{li:two}.

A 90\% confidence level upper limit on the branching fraction of the $\taupeta$ decay 
is measured to be  $\CL$.
This is an order of magnitude lower than the previous $90\%$ confidence level upper limit of $7.4 \times 10^{-5}$ set by the CLEO collaboration \cite{cleo:fpi}.
No significant evidence for this second-class current decay mode of the \mtau is observed.

We are grateful for the 
extraordinary contributions of our \pep2\ colleagues in
achieving the excellent luminosity and machine conditions
that have made this work possible.
The success of this project also relies critically on the 
expertise and dedication of the computing organizations that 
support \babar.
The collaborating institutions wish to thank 
SLAC for its support and the kind hospitality extended to them. 
This work is supported by the
US Department of Energy
and National Science Foundation, the
Natural Sciences and Engineering Research Council (Canada),
the Commissariat \`a l'Energie Atomique and
Institut National de Physique Nucl\'eaire et de Physique des Particules
(France), the
Bundesministerium f\"ur Bildung und Forschung and
Deutsche Forschungsgemeinschaft
(Germany), the
Istituto Nazionale di Fisica Nucleare (Italy),
the Foundation for Fundamental Research on Matter (The Netherlands),
the Research Council of Norway, the
Ministry of Science and Technology of the Russian Federation, 
Ministerio de Educaci\'on y Ciencia (Spain), and the
Science and Technology Facilities Council (United Kingdom).
Individuals have received support from 
the Marie-Curie IEF program (European Union) and
the A. P. Sloan Foundation.


\begin{thebibliography}{99}
\bibitem{chconj}
Charge conjugation is assumed throughout this paper.
\bibitem{novo}
The Novosibirsk function is defined as \\
$f(m) = A \; {\rm exp} ( -0.5 \left\{ {\rm ln}^2 \left[ 1 + \Lambda \tau \cdot
(m - m_0) \right] / \tau^2 + \tau^2 \right\} )$,
where $\Lambda = \sinh(\tau \sqrt{{\rm ln}4}) / (\sigma \tau \sqrt{{\rm ln}4}
)$,
the peak position is $m_0$, the width is $\sigma$ and
$\tau$ is the tail parameter.
\bibitem{pvals}
The functional form of the P-wave Breit-Wigner function is identical to the nominal Breit-Wigner function except that the width
$\Gamma_{f_1}$ is modified by a $(p(x)/p(0))^3$ factor. The momentum terms $p(x)$ and $p(0)$ are the momenta of the outgoing
particles in a two-body decay ($f_{1}\rightarrow\pi^-a_0^+(980)$) evaluated at the current ($x$) and peak ($m_{f_1}$) mass values.
\end{thebibliography}
\end{document}